\documentclass[aps,prd,onecolumn,amssymb]{revtex4}
\usepackage{graphicx,bm,color}
\usepackage[english]{babel}
\usepackage{amsmath}
\usepackage{amsfonts}
\newcommand{\beq}{\begin{eqnarray}}
\newcommand{\eeq}{\end{eqnarray}}

\def\nn{\nonumber}
\def\bce{\begin{center}}
\def\ece{\end{center}}
\usepackage{enumitem}
\begin{document}

\title{Exact Solutions in Modified Gravity Models}
\author{ Andrey N. Makarenko  and  Valery V. Obukhov}

\affiliation{Tomsk State Pedagogical University, 634061,  Kievskaya st., 60,
Tomsk, Russia}

\begin{abstract}
We review the exact solutions in modified gravity. It is
one of the main problems of mathematical physics for the gravity
theory. One can obtain an exact solution if the field equations
reduce to a system of ordinary differential equations. In this paper
we consider a number of exact solutions obtained by the method of
separation of variables. Some applications to Cosmology and BH
entropy are briefly mentioned.
\end{abstract}

\maketitle

%%%%%%%%%%%%%%%%%%%%%%%%%%%%%%%%%%%%%%%%%%%%%%%%%%%%%%%%%%%%

\section{Introduction}

Gravitational field equations describing the geometry of space-time
play a fundamental role in modern theoretical physics. Their
analysis is an extremely difficult task. However, one can find an
exact solution in some cases, imposing certain additional
restrictions. There are several ways to impose constraints on the
space-time, for example, the algebraic classification of the Weyl
tensor (of Petrov types) and the Ricci tensor (Plebanski type), the
choice of energy-momentum tensor from physical considerations,
presence of the symmetry groups acting on a manifold,~\textit{etc}.

One can get the exact solution by reducing the system of equations
to a system of ordinary differential equations. This can be done
using the method of separation of variables. In essence, the
separation of variables is the only currently known method of
structural integration of the field equations. The purpose of the
method consists of classification of all the privileged coordinate
systems and external fields, which is a separation of variables. In
the classification we refer to the transfer of all relevant
space-time metrics (non-equivalent with respect to admissible
coordinate transformations) satisfying the requirement of complete
separation of variables in the equations of motion of test
particles. In flat space-time classification is carried out
completely.

From a mathematical point of view,  the study of homogeneous spaces
and St\"{a}ckel metrics is of special interest. The spaces are
united by existence in the space of sets consisting of three
geometric objects (for St\"{a}ckel spaces---Killing vector and
tensor fields, for homogeneous spaces---Killing vector fields). For
both cases the field equations can be reduced to a system of finite
(but sufficiently large) number of ordinary differential equations.
There are many methods in mathematical physics for studying such
systems of equations. For example, one can use methods of additional
symmetries of the system of equations, the Hamiltonian formulation
of the theory of dynamical systems,~\textit{etc}. Essentially, all
physically interesting cases (FRW cosmology, BH) belong to
St\"{a}ckel and homogeneous spaces.
%%%%%%%%%%%%%%%%%%%%%%%%%%%%%%%%%%%%%%%%%%%%%%%%%%%%%%%%%%%%

\section{Exact Solutions in St\"{a}cke and Homogeneous Spaces}

Recall that metric is called the St\"{a}ckel one if the
Hamilton--Jacobi equation
\begin{equation}
g^{\alpha\beta}S_{,\alpha}S_{,\beta}=m^2 \qquad \alpha,
\beta=1,\ldots,n \label{s1}
\end{equation}
can be integrated by the method of complete separation of variables.
The privileged co-ordinate set ${u^\alpha}$ exists for
which complete integral of Equation (\ref{s1}) can be shown in the
form
\begin{equation}
S=\sum_{i=1}^n \phi_i(u^i,\lambda) \label{s2}
\end{equation}
where $\lambda_i$  is the essential parameter.

It appears that the other important equations of motion
(Klein-Gordon-Fock, Dirac, Weyl) can be integrated by complete
separation of variables  only for the metrics belonging to the class
of St\"{a}ckel spaces.

That is why the research of this class of spaces belongs to the one
of the important problems of the mathematical physics.

The metrics of the St\"{a}ckel spaces can be used for integrating
the field equations of General Relativity (GR) and other theories of
Gravity. Note that such famous solutions as Schwarzschild, Kerr,
NUT, Friedman and others belong to the class of St\"{a}ckel spaces.
Apparently the first papers devoted to the problem of classification
of the St\"{a}ckel spaces satisfying the Einstein equations were
published by Carter \cite{b1}. Later in our paper \cite{b2} the
complete classification of the special St\"{a}ckel electrovacuum
spaces has been found. In other words, all St\"{a}ckel spaces
satisfying the Einstein-Maxwell equations for the case when
potentials $A_i$ admit complete separation of variables for
Hamilton--Jacobi equation have been found. In our paper \cite{b3}
the classification problem has been solved for the case when $A_i$
are arbitrary functions and spaces are null (types N.1). In our
paper \cite{b4} all electrovacuum spacetimes admitting
diagonalization and complete separation of variables for the
Dirac--Fock--Ivanenko equation were found.

One of the complicated problems of the modern mathematical physics
is the integration problem of the Einstein-Dirac equations.

Using spaces for which Equation (\ref{s1}) can be integrated by the
complete separation of variables and separated solutions of the
Dirac equation, one can transform Einstein-Dirac equations to the
set of functional equations. The first papers devoted to the
classification problem for the Einstein-Dirac equations were done
by Bagrov, Obukhov, Sakhapov \cite{b5}. The St\"{a}ckel spaces of
type (3.1) for Einstein-Dirac and Einstein-Weyl equations have
been studied. Appropriate solutions have been obtained. They contain
arbitrary functions depending on null variable only.

The problem of classification of St\"{a}ckel spaces for other
theories of gravity for the first time was considered in
papers \cite{b6}-\cite{b63}.

We have solved the classification problem for the Einstein-Vaidya
equations. Let the stress-energy tensor have the form
\begin{equation}
T_{\alpha\beta}= a(x)l_\alpha l_\beta,\qquad l_\alpha l^\alpha=0
\label{s3}
\end{equation}

The solution of this problems, as well as a detailed overview of the
theory of St\"{a}ckel spaces, can be found in~\cite{OB1}.

Of high interest are homogeneous spaces, which lie at the heart of
modern cosmology~\cite{lit6}-\cite{lit63}. The homogeneous spaces are a base for
building the Big Bang model, the initial singularity, as well as the
inflationary model. It is of interest to identify the various
mechanisms of isotropization of the universe~\cite{lit10}-\cite{lit105}.
Homogeneous spaces are also used in a variety of modern theories of
gravity for the study of general regularities in the picture of the
universe~\cite{lit17}-\cite{lit173}. One can study effects of the gravitational
field to other fields and matter on the background of homogeneous
spaces~\cite{lit23}-\cite{lit234}.

Let us consider the Einstein--Weyl equations. One can show that for
all types of Bianchi classification, a closed self-consistent system
of ordinary integrable differential equations can be
constructed~\cite{AN}-\cite{AN111}. For example, consider the first type of
classification of Bianchi~\cite{AN1}. This is one of the most simple
cases; to construct a general solution for all types is not
possible.

All calculations will be carried out in the Newman--Penrose
formalism. Einstein equation takes the~form
\begin{equation}
\left\{\begin{aligned}
\Phi_{00}&=G T_{000'0'}\\
\Phi_{01}&=G T_{000'1'}\\
\Phi_{02}&=G T_{001'1'}\\
\Phi_{11}&=G T_{010'1'}+\frac{G}{4}T^{AB'}_{AB'}\\
\Phi_{12}&=G T_{011'1'}\\
\Phi_{22}&=G T_{111'1'}\\
\Lambda&=\frac{1}{6} H-\frac{G}{12} T^{AB'}_{AB'}
\end{aligned}\right.
\end{equation}
Here $H$ is cosmological constant, $G$ is corresponds to the
gravitational constant, $\Phi_{ab}$ are Ricci spinors and
$T_{ABA'B'}$ is energy-momentum tensor,
\begin{align}
\label{Tenz1}
\begin{split}
T_{AB'CD'}=ik(\xi_{D'}\nabla_{AB'}\xi_C + \xi_B\nabla_{CD'}\xi_A &-
\xi_C\nabla_{AB'}\xi_{D'} - \xi_A\nabla_{CD'}\xi_{B'}-
\eta_{D'}\nabla_{AB'}\eta_C   \\
&-\eta_B\nabla_{CD'}\eta_A + \eta_C\nabla_{AB'}\eta_{D'} +
\eta_A\nabla_{CD'}\eta_{B'})
\end{split}
\end{align}
and $\nabla_{AB}$ is spinor derivative.

Ricci spinors are expressed in terms of spin factors  as follows
\begin{align*}
\Phi_{00}&=D\rho-\overline{\delta}\kappa
-\rho^2-\sigma\overline{\sigma}-\rho(\varepsilon+\overline{\varepsilon})+
\overline{\kappa}\tau+\kappa(3\alpha+\overline{\beta}-\pi),
\\
\Phi_{01}&=D\overline{\alpha}-\delta\overline{\varepsilon}
-\overline{\alpha}(\overline{\rho}+\varepsilon-2\overline{\varepsilon})
-\overline{\beta}\sigma+\beta\overline{\varepsilon}+\overline{\kappa}
\overline{\lambda}+\kappa\overline{\gamma}-
\overline{\pi}(\overline{\varepsilon}+\overline{\rho}),
\\
\Phi_{02}&= D\overline{\lambda}-\delta\overline{\pi}
+\overline{\lambda}(3\overline{\varepsilon}-\varepsilon-\overline{\rho})
-\overline{\mu}\sigma-\overline{\pi}(\overline{\pi}-\beta+\overline{\lambda})
+\overline{\nu}\kappa,
\\
\Phi_{11}&=\frac{1}{2}[D\gamma-\Delta\varepsilon+\delta\alpha-\overline{\delta}
\beta-\alpha(\tau+\overline{\pi})-\beta(\overline{\tau}+\pi)+
\gamma(\varepsilon+\overline{\varepsilon})
\\
&\quad+\varepsilon(\gamma+
\overline{\gamma})-\tau\pi+\kappa\nu-\mu\rho+
\sigma\lambda-\alpha\overline{\alpha}-\beta\overline{\beta}+2\alpha
\overline{\beta}-
\\
&=\gamma(\rho-\overline{\rho})-\varepsilon(\mu-\overline{\mu})],
\\
\Phi_{12}&=\delta\gamma-\Delta\beta
-\gamma(\tau-\overline{\alpha}-\beta)-\mu\tau+\sigma\nu+\varepsilon\overline{\nu}
+\beta(\gamma-\overline{\gamma}-\mu)-\alpha\overline{\lambda},
\\
\Phi_{22}&=\delta\nu-\Delta\mu
-\mu^2-\lambda\overline{\lambda}-\mu(\gamma+\overline{\gamma})+\overline{\nu}\pi
-\nu(\tau-3\beta-\overline{\alpha}),
\\
6\Lambda&=2[\overline{\delta}\tau-\Delta\rho-\rho\overline{\mu}
-\sigma\lambda+\tau(\overline{\beta}-\alpha-\overline{\tau})+\rho(\gamma
+\overline{\gamma}+\nu\kappa]-
\\
&=D\gamma+\Delta\varepsilon+\delta\alpha-\overline{\beta}+\alpha(\tau+
\overline{\pi})+\beta(\overline{\tau}+\pi)-\gamma(\varepsilon+\overline{\varepsilon})-
\\
&=\varepsilon(\gamma+\overline{\gamma})+\tau\pi-\kappa\nu-\mu\rho+
\lambda\sigma
\end{align*}

Choose a metric in the form
\[
g_{00}=1,\;\;\; g_{0\alpha}=0,\;\;\; g_{ij}=-\gamma_{ij}
\]
where $\gamma_{ij}$ is the metric of a three-dimensional space with
the signature $(+,+,+)$. It is simple to establish that this space
permits a three-parameter Abelian group of motions, and hence is a
Steckel space of type~(3.0).

The orthogonal tetrad is chosen in the form
\begin{align*}
\begin{aligned}
e_{(0)\alpha}&=(1,0,0,0),   &e_{(1)\alpha}&=(0,A,B,C)\\
e_{(2)\alpha}&=(0,K,S,V), &e_{(3)\alpha}&=(0,P,M,Z)
\end{aligned}
\end{align*}
where $A,B,C,R,S,V,P,M,Z$  are arbitrary functions of the time.

Using this tetrad, we construct the Newman--Penrose tetrad
\begin{align*}
\begin{aligned}
\ell_i&=\frac{1}{\sqrt{2}}(e_{(0)i}+e_{(1)i}), &
n_i&=\frac{1}{\sqrt{2}}(e_{(0)i}-e_{(1)i}),\\
m_i&=\frac{1}{\sqrt{2}}(e_{(2)i}+i e_{(3)i}), &
\overline{m_i}&=\frac{1}{\sqrt{2}}(e_{(2)i}-i e_{(3)i})
\end{aligned}
\end{align*}

and obtain the following relations between the spin factors
\begin{align*}
\begin{aligned}
\lambda&=-\overline{\sigma}, &\nu&=\overline{\kappa},\;\;
\pi=\overline{\tau}, &\gamma&=-\overline{\varepsilon}\\
\alpha&=\overline{\beta}, &\mu&=-\rho=\overline{\mu},
&\alpha&=\frac{1}{2}(\overline{\tau}-\overline{\kappa})
\end{aligned}
\end{align*}

The spinor field of spatial rotation of the tetrad may be
diagonalized and made real (one real component remains). For this
case, the energy-momentum tensor takes the form (\ref{Tenz1}).
\begin{align*}
\begin{aligned}
T_{00'00'} & =  2 i k a \xi_0^2
(\overline{\varepsilon}-\varepsilon), &T_{00'01'}  &=  i k a \xi_0^2
\tau,
&T_{01'01'} &= - 2 i k a \xi_0^2\sigma\\
T_{01'11'} & =   i k a \xi_0^2 \kappa, &T_{11'11'}  &=  0,
&T_{00'11'}  &=   i k a \xi_0^2
(\overline{\varepsilon}-\varepsilon), \;\; T_{01'10'} = 0
\end{aligned}
\end{align*}
The field equations take the form
\begin{align}
\label{n3.10}
\dot{\rho}-\rho^2-\sigma\overline{\sigma}-2\rho\varepsilon
         -4\kappa\overline{\kappa}& = 0\\
\label{n3.11}
\dot{\sigma}-2\rho\sigma+2\sigma\varepsilon& = 2iy{\xi_0}^2\sigma\\
\label{n3.12}
\dot{\kappa}-2\rho\kappa+2\kappa\varepsilon& = iy{\xi_0}^2\kappa\\
\label{n3.13}
2\dot{\varepsilon}-\dot{\rho}+\rho^2+\sigma\overline{\sigma}-2\rho\varepsilon
         -4\kappa\overline{\kappa}+4\varepsilon^2& = \frac{H}{2}\\
\label{n3.14} \dot{\rho}-1\rho^2+2\rho\varepsilon& = \frac{H}{2}\\
\label{n3.15} \dot{\xi_0}+(\varepsilon-\rho)\xi_0& = 0
\end{align}
where $\dot{ }=\frac{d}{dt}$.

Multiplying Equation~(\ref{n3.11}) by $\overline{\sigma}$ and
Equation~(\ref{n3.12}) by $\overline{\kappa}$, we add and subtract
the resulting equations and their conjugate forms. Transformation
yields
\begin{equation}
\label{n3.17}
\begin{split}
\dot{(ln\kappa\overline{\kappa})}&=-4(\varepsilon-\rho)\\
\dot{(ln\sigma\overline{\sigma})}&=-4(\varepsilon-\rho)
\end{split}
\end{equation}

Multiplying Equation~(\ref{n3.15}) by $y\xi_0$, we obtain
\begin{equation}
\label{n3.19} \dot{\left[\ln \left( y{\xi_0}^2
\right)\right]}=-2(\varepsilon-\rho)
\end{equation}

One find that
\begin{equation}
\begin{split}
\kappa=k y \xi_0^2e^{i\omega_1}\\
\sigma=s y \xi_0^2e^{i\omega_2}
\end{split}
\end{equation}

Finally, we obtain the following equation
\begin{equation}
\label{n3.21}
\dot{(\varepsilon-\rho)}+2(\varepsilon-\rho)^2=\frac{3H}{4}
\end{equation}

(1) Consider the case when $H=0$. In this case
\begin{equation}
\frac{\dot{\left(\varepsilon-\rho\right)}}
{\left(\varepsilon-\rho\right)^2}=-2 \nn
\end{equation}
and thus
\begin{equation}
\frac{1}{\left(\varepsilon-\rho\right)}=2x+const \nn
\end{equation}
Incorporating the constant into the definition of $x$, we write
\begin{equation}
\label{n3.22} \left(\varepsilon-\rho\right)=\frac{1}{2x}
\end{equation}
Hence
\begin{equation}
\label{n3.23}
\frac{1}{2x}=-\frac{1}{2}\frac{\ddot{\omega}}{\dot{\omega}}\;\;\;\;
\Rightarrow\;\;\;\;\dot{\omega}=\frac{d}{x}
\end{equation}
where $d$ is a constant of integration. We now find P and e in
explicit form
\begin{align}
\label{n3.24} \dot{\rho}+\frac{\rho}{x}=0
&\Rightarrow\rho=\frac{f}{x} \\
%\end{equation}
%\begin{equation}
\label{n3.25} \varepsilon-\rho=\frac{1}{2x}
&\Rightarrow\varepsilon=\frac{1}{2x}+\frac{f}{x}=\frac{f+1/2}{x}
\end{align}
where $f$ is a constant of integration. The constants of integration
satisfy the condition
\begin{equation}
2f+3f^2+d^2({s}^2+4{k}^2)=0 \nn
\end{equation}
Obviously, this equation has a nontrivial solution.

(2) Similarly, for the case $H \ne 0$, we can show that it is not
implemented.

To find the tetrad we have to solve the two equations
\begin{align}
2A\varepsilon - (K-iP)\kappa -(K+iP)\overline{\kappa}&=A'\label{n16}\\
2A\kappa + (K-iP)\sigma +(K+iP)\rho &=\hbox{}-(K' +i P')\label{n17}
\end{align}
Since
\begin{equation}
\varepsilon= \frac{1/2+f}{x},\;\;\;\; \rho=\frac{f}{x},\;\;\;\;
\kappa=k\frac{d}{x}e^{i\omega},\;\;\;\;
\sigma=s\frac{d}{x}e^{i(2\omega+c)} \nn
\end{equation}
they take the form
\begin{align*}
\begin{split}
2A\frac{1/2+f}{x}-(K-iP)\frac{kd}{x}e^{i\omega}-(K+iP)\frac{kd}{x}e^{-i\omega}&=A'\\
2A\frac{kd}{x}e^{i\omega}+(K-iP)\frac{sd}{x}e^{i(2\omega+c)}
(K+iP)\frac{f}{x}&=\hbox{}-(K' +i P')
\end{split}
\end{align*}

We multiply the second equation by $e^{-i\omega}$ and make the
substitution
\begin{equation}
\left\{
\begin{array}{@{}c}
X=K\cos \omega +P \sin \omega\\
Y=P\cos \omega -K \sin \omega
\end{array}
\right.
 \nn
\end{equation}
The inverse transformations are
\begin{equation}
\left\{
\begin{array}{@{}c}
K=Y\cos \omega +X \sin \omega\\
P=X\cos \omega -Y \sin \omega
\end{array}
\right.
 \nn
\end{equation}

We also introduce the notation $\frac{f}{d}=F, \frac{1}{d}=D$
\[
\begin{aligned}
A(2F+D)-k(X-iY)-k(X+iY)&=\frac{A'}{\omega '}\\
2Ak+s(X-iY)e^{ic}+F(X+iY)&=\frac{\hbox{}-\left(X+iY\right)'-i\omega
'(X+iY)}{\omega '}
\end{aligned}
\]

We make one more substitution
\begin{equation}
A=We^{-F\omega},\;\;\;\; (X+iY)=(U+iV)e^{-F\omega}
 \nn
\end{equation}
and take into account that
\[
\frac{A'}{\omega '}=\frac{dA}{dx}\frac{dx }{d\omega }=\frac{dA}{d
\omega}
\]

Then the equations take the form
\[
\begin{aligned}
W(3F+D)-k\left[(U+iV)+(U-iV) \right]&=\frac{d W}{d\omega}\\
2kW+se^{ic}(U-iV)&=\hbox{}-\frac{d}{d\omega}(U+iV)-i(U+iV)
\end{aligned}
\]

If $e^{ic}=\alpha+i\beta$, $l=3F+D$ and $\alpha^2+\beta^2=1$, then
\[\left\{
\begin{aligned}
Wl-2kU&=\frac{dW}{d\omega}\\
2kW+c_2\alpha U+ (s\beta -1)V&=\hbox{}-\frac{dU}{d\omega}\\
(s\beta +1)U-s\alpha V&=\hbox{}-\frac{dV}{d\omega}
\end{aligned}
\right.
\]

One can obtain a third-order equation in V
\begin{equation}
\frac{d^3V}{d\omega^3}-\frac{d^2V}{d\omega^2}l-\frac{dV}{d\omega}
(s^2+4k^2-1)+V(4k^2s\alpha+l(s^2-1))=0
 \nn
\end{equation}
The characteristic equation takes the form
\begin{equation}
\label{n3.28} \lambda^3-\lambda^2l-\lambda
(s^2+4k^2-1)+(4k^2s\alpha+l(s^2-1))=0
\end{equation}

From this it is possible to find three roots $\lambda_1$,
$\lambda_2$, $\lambda_3$
\begin{equation}
\begin{array}{r@{\;} l}
\lambda^3-\lambda^2l-\lambda (s^2+4k^2-1)+(4k^2s\alpha+l(s^2-1))&=\\
\lambda^3-\lambda^2(\lambda_1+\lambda_2+\lambda_3)
-\lambda(\lambda_1\lambda_2+\lambda_1\lambda_3+\lambda_2\lambda_3)
-\lambda_1\lambda_2\lambda_3&=(4k^2s\alpha+l(s^2-1))
\end{array}
 \nn
\end{equation}

\begin{equation}
\left\{
\begin{aligned}
\lambda_1+\lambda_2+\lambda_3&=l\\
\lambda_1\lambda_2+\lambda_1\lambda_3+\lambda_2\lambda_3&=\hbox{}-(s^2+4k^2-1)\\
\lambda_1\lambda_2\lambda_3&=\hbox{}-(4k^2s\alpha+l(s^2-1))
\end{aligned}
\right.
 \nn
\end{equation}
In all, four cases are possible:

\begin{enumerate}
\item $\lambda_1$, $\lambda_2$,
$\lambda_3$ are pairwise unequal and real. Then
\begin{align*}
V=c_1e^{\lambda_1 \omega}+c_2e^{\lambda_2 \omega}+c_3e^{\lambda_3
\omega}
 \nn
\end{align*}

\item $\lambda_1$ is real, $\lambda_2=\overline{\lambda_3}$
\begin{align*}
V=c_1e^{\lambda_1 \omega}+c_2e^{\lambda_2
\omega}+c_3e^{\overline{\lambda_2} \omega}
 \nn
\end{align*}

\item $\lambda_1$, $\lambda_2=\lambda_3$ are real
\begin{align*}
V=c_1 e^{\lambda_1 \omega}+(c_2+c_3\omega)e^{\lambda_2 \omega}
 \nn
\end{align*}

\item $\lambda_1=\lambda_2=\lambda_3=\lambda$
\begin{align*}
\begin{aligned}
V&=(c_1+c_2\omega+c_3\omega^2)e^{\lambda \omega}
\\
%\end{align*}
%
%\begin{align*}
U &= \frac{-\frac{dV}{d\omega}+b\alpha V} {b\beta+1}\\
W &= \frac{\frac{d^2V}{d\omega^2}-(b^2-1)V}{2a(b\beta+1)}
\end{aligned}
\end{align*}
\end{enumerate}

Thus, an accurate solution of the Einstein--Weyl equation has been
obtained for space-time of Bianchi type 1. The solution only exists
when $H = 0$. The spin coefficients and the Weyl spinor take the
form
\begin{equation}
\rho=\frac{f}{x},\;\;\;\; \varepsilon=\frac{1/2+f}{x},\;\;\;\;
\sigma=\frac{s d}{x}e^{i(2\omega+c)},\;\;\;\; \kappa=\frac{k
d}{x}e^{i\omega},\;\;\;\; y{\xi_0}^2=\frac{d}{x}
 \nn
\end{equation}
where $f, k, s, y, c$  are constants and $\omega=d \ln x$

On integrating the equations for the tetrad, the functions obtained
will depend on the form of Equation~(\ref{n3.28}). As an example,
the functions for the first case, when all the roots are different
and real, are
\begin{align*}
\begin{aligned}
A&=e^{-F\omega}
\left[\frac{c_1e^{\lambda_1\omega}({\lambda_1}^2-b^2+1)+
            c_2e^{\lambda_2\omega}({\lambda_2}^2-b^2+1)+
            c_3e^{\lambda_3\omega}({\lambda_3}^2-b^2+1)}{2a(b\beta+1)}
\right]\\
K&=e^{-F\omega}
\left[c_1e^{\lambda_1\omega}\frac{(b\alpha-\lambda_1)}{b\beta+1}+
      c_2e^{\lambda_2\omega}\frac{(b\alpha-\lambda_2)}{b\beta+1}
+c_3e^{\lambda_3\omega}\frac{(b\alpha-\lambda_3)}{b\beta+1}
\right]\\
P&=e^{-F\omega}
\left[c_1e^{\lambda_1\omega}+c_2e^{\lambda_2\omega}+c_3e^{\lambda_3\omega}
\right]
\end{aligned}
\end{align*}
where $\lambda_1$, $\lambda_2$, $\lambda_3$ are the roots of
Equation~(\ref{n3.28}); $F, a, b, c_1, c_2, c_3, \alpha, \beta$ are
constant.

The other functions specifying the tetrad are of analogous form. The
only difference is that another set of arbitrary constants must be
chosen in place of $ c_1, c_2, c_3$.

\section{Modified Gravity}

In the previous sections a few exact solutions of the classical
theory of gravitation were obtained. However, in recent years,
modified gravity theory is very popular. These theories my be
studied also using the methods described above. Typically, solutions
with a spatially flat metric depend only on time. It corresponds to
the first type of Bianchi and type (3.1) on the classification of
St\"{a}ckel.

The most popular models are models of modified
gravity~\cite{N1} - \cite{N111}, which represents a classical generalization
of general relativity (modifications of the Hilbert--Einstein action
by introducing different functions of the Ricci scalar~\cite{N11} - \cite{fr}
or Gauss--Bonnet invariant~\cite{gb}-\cite{gb55}), should consistently
describe the early-time inflation and late-time acceleration,
without the introduction of any other dark component.

In the framework of these theories, a number of cosmological models
have been constructed to adequately describe the current
observational data~\cite{Dat}-\cite{Dat5}. In addition, under this theory there
may be objects such as black holes. The properties of these objects
are different from classical ones. For example the black hole
entropy in the model F (R) gravity will have the form~\cite{zer, zer1}
\begin{equation}
S=\frac{A_H}{4} f'(R_0)
\end{equation}
where $A_H=4\pi r_H^2$.

As an example of the exact solution in modified theory of gravity.
We consider the sixth dimensional Gauss--Bonnet theory~\cite{mak1, mak11}.

We shall start from the following string-inspired action in six
dimensions
\[
S=\int d^6 x\sqrt{-g} \left(R+\epsilon L_{GB}\right)
\]
where $\epsilon$ is a constant and $ L_{GB}$ is Gauss--Bonnet
invariant
\begin{equation}
L_{GB}= R_{\mu \nu \alpha \beta} R^{\nu \mu \alpha \beta} - 4 R_{\mu
\nu} R^{\mu \nu} + R^2
\end{equation}
The metric is the product of the usual metric corresponding to the
4-dimensional FRW universe and a 2-dimensional surface, namely 
\begin{equation}
ds^2=-dt^2 + a^2(t) \left[(dx^1)^2+(dx^2)^2+(dx^3)^2\right]+b^2(t) \left[(dx^4)^2+(dx^5)^2\right],
\end{equation}
the scalar curvature is 
\begin{equation}
R=\frac{6 {\dot a}^2}{a^2}+\frac{12 \dot 
a\dot b}{ab}+\frac{2{\dot b}^2}{b^2}+\frac{6 \ddot a}{a}+\frac{4 \ddot b} {b},
\end{equation}
while the four-dimensional and topologically invariant Gauss--Bonnet
Lagrangian, $L_{GB}$, has the form 
\begin{equation}
 L_{GB}=\frac{48 {\dot
a}^3\dot b}{a^3 b}+\frac{72 {\dot a}^2{\dot b}^2}{a^2 b^2}+\frac{24 {\dot a}^2\ddot a}{a^3}+\frac{96 \dot a \ddot a \dot b}{a^2 b}+
\frac{24 \ddot a{\dot b}^2}{ab^2} +\frac{48 {\dot a}^2\ddot b}{a^2 b}+  \frac{48 \dot a\dot b\ddot b}{a b^2},
\end{equation}
or, equivalently,
\begin{equation}
L_{GB}=\frac{24}{a^3 b^2}(2{\dot a}^3 b\dot b+3 a {\dot a}^2 {\dot b}^2+{\dot a}^2 \ddot a b^2+4 a \dot a\ddot a b\dot b+  a^2\ddot a {\dot b}^2+
2 a{\dot a}^2 b\ddot b+2 a^2\dot a\dot b\ddot b).
\end{equation}

The corresponding equations of motion are obtained by variation of
the action with respect to $a$ and $b$, which yields
\begin{eqnarray}
{\dot a}^2 b^2+ 4 a\dot  a b\dot b+ a^2{\dot b}^2+ 2 a \ddot a b^2+2 a^2 b\ddot b+12 \epsilon {\dot a}^2 {\dot b}^2+ 16\epsilon \dot a\ddot a b\dot b+8 \epsilon a\ddot a {\dot b}^2     +  8\epsilon {\dot a}^2 b \ddot b+16 \epsilon a \dot a \dot b \ddot b&=0, \nonumber \\
     3 a{\dot a}^2 b+ 3 a^2 \dot a \dot b+ 3 a^2 \ddot a b+a^3 \ddot b+12 \epsilon {\dot a}^3 \dot b+12 \epsilon {\dot a}^2 \ddot a b+ 24 \epsilon a \dot a \ddot a \dot b+12 \epsilon a {\dot a}^2 \ddot b&=0. \label{2a}
\end{eqnarray}
These equations can be easily rewritten in terms of the Hubble rates
$H=a'/a$ and $h=b'/b$, namely

\begin{equation}
 \begin{split}
  3 h^2+4 h H+3 H^2+2 \dot h+2 \dot H+16 \epsilon h^3 H+28 \epsilon h^2 H^2+16 \epsilon h H^3
 +16 \epsilon h \dot h H+8 \epsilon \dot h H^2+8 \epsilon h^2 \dot H+16 \epsilon h H \dot H&=0,  \\
 h^2+3  h H+6 H^2+\dot h+3 \dot H+12 \epsilon h^2 H^2+36 \epsilon h H^3+12 \epsilon H^4 
+12 \epsilon \dot h H^2+24\epsilon hH\dot H+12\epsilon H^2\dot
H&=0.\label{2b}
\end{split}
\end{equation}
In addition, variation over the metric in the above expressions
gives the constraint equation 
\begin{equation}
 h^2+ 6 h H+3 H^2+36 \epsilon h^2 H^2+24 \epsilon h H^3=0 \label{connect}.
\end{equation}

 This equation helps to
exclude $h$ and $h'$ from Equation (\ref{2b}). As a result, one gets
an equation for $H$ only:
\begin{equation}
 H'= 3 H^2 \times \frac{\sqrt 6+4 G+\epsilon (-22 \sqrt 6+64 G)H^2-24 \epsilon ^2 (9 \sqrt 6-52G) H^4+96 \epsilon ^3 (17 \sqrt 6+12 G) H^6-8064 \sqrt 6 \epsilon ^4 H^8}
{\sqrt 6-12 \epsilon (\sqrt 6+16 G) H^2+72 \epsilon^2 (3 \sqrt 6-32 G) H^4-2880 \sqrt 6 \epsilon^3 H^6+31104 \sqrt 6 \epsilon ^4 H^8},
\end{equation}
where
\begin{equation} 
G=\sqrt{1-6\epsilon H^2+24 \epsilon^2H^4}\end{equation}
One can
check that this last equation obeys the fundamental relation (for
$\epsilon >0$): 
\begin{equation} 
H'=\frac{H^2(H^2-p^2)}{(H^2-q^2)(H^2-r^2)}f(H)
\end{equation}
where $p, q$ and
$r$ are constants, and the function \ $f(H)<0$.

For $\epsilon =0$  one does recover (as it should be) an explicit
solution. In other cases, only a numerical analysis can be carried
out. In fact, the equations for $a, b$ are
\begin{equation}
b^2 {a'}^2+4  a b a' b'+a^2 {b'}^2+2 a b^2 a''+2 a^2 b b''=0, \;\;
3 a b{a'}^2+3 a^2 a' b'+3 a^2 b a''+a^3 b''=0
\end{equation}
and, from here,
\begin{equation}
\begin{split}
3 h^2+4 h H+3 H^2+2 h'+2 H'&=0  \\
h^2+3  hH+6 H^2+h'+3 H'&=0
\end{split}
\end{equation}
from where one gets that $ H'=(3\pm 2\sqrt 6)H^2$ and the
solution is given by $ H=-\frac 1{\alpha t+C_1}$ being $\alpha=3\pm 2\sqrt 6$. Moreover, in terms of the scale factors:
\begin{equation}
\begin{split}
 a&=C_2[\pm (\alpha t+C_1)]^{-1/\alpha}  \\
b&=C_3[\pm (\alpha t+C_1)]^{\beta/\alpha}
\end{split}
\end{equation}
where $ \beta=3\pm \sqrt 6$.
%\ms
General approach to St\"{a}ckel spaces of first sections may be
applied to such theory as well. However, this is more complicated in
modified gravity.

\section{Conclusions}

In this article we consider the exact solutions constructed in the
classical theory of St\"{a}ckel spaces  and partially homogeneous
spaces. We consider in detail the exact solution of Einstein--Weyl
space of the first type of classification of Bianchi in the
Newman--Penrose formalism. As an example we consider the exact
solution for the modified gravity theory of the Gauss--Bonnet for
the six-dimensional  metric which depends only on time and has a
diagonal form. Thus, we discuss the problem of obtaining exact
solutions in the different theories of gravity. The paper shows how,
by use of the method of separation of variables, can one construct
exact solutions of cosmological models, both for space with matter
(the classical theory of gravity) and in the case of the modified
theories of gravity (six-dimensional theory of the Gauss--Bonnet).

Note that separation of field equations for modified gravity in
St\"{a}ckel space may be done in analogy with the method developed
at the beginning of this work. However, the corresponding
generalization is very cumbersome technically.

\end{document}